\documentclass[preprintnumbers, floatfix,
preprintnumbers, letterpaper, superscriptaddress,nofootinbib, twocolumn]{revtex4}

\usepackage{graphicx}
\usepackage{microtype}
\usepackage{amsmath}
\usepackage{amssymb}
\usepackage{subfigure}
\usepackage{hyperref}
\usepackage{url}
\usepackage{xcolor}
\usepackage{color}
\usepackage{mathrsfs}
\usepackage{calrsfs}
\usepackage{amsfonts}
\usepackage{latexsym}
\usepackage{ragged2e}
\usepackage{epsfig}
\usepackage{textcomp}

\usepackage{textcomp}
\usepackage{phaistos}

\usepackage{caption}
\DeclareCaptionJustification{justified}{\leftskip=0pt \rightskip=0pt \parfillskip=0pt plus 1fil}
\definecolor{vividviolet}{rgb}{0.62, 0.0, 1.0}
\definecolor{amaranth}{rgb}{0.9, 0.17, 0.31}
\definecolor{palatinateblue}{rgb}{0.15, 0.23, 0.89}
\definecolor{brightpink}{rgb}{1.0, 0.0, 0.5}
\definecolor{cornflowerblue}{rgb}{0.39, 0.58, 0.93}
\definecolor{deepcarminepink}{rgb}{0.94, 0.19, 0.22}
\definecolor{radicalred}{rgb}{1.0, 0.21, 0.37}

\hypersetup{ linktoc=all,
    colorlinks, linkcolor={palatinateblue},
    citecolor={brightpink}, urlcolor={amaranth}
}

\graphicspath{{Images/}}

\makeatletter
\def\@fnsymbol#1{{\ifcase#1 \or \PHrosette \or \PHplaneTree  \or \textleaf  \else\@ctrerr\fi}}
	\makeatother

\begin{document}

\title{A Note on Smarr Relation and Coupling Constants}

\author{Shi-Qian \surname{Hu}}
\email{mx120170256@yzu.edu.cn}
\affiliation{Center for Gravitation and Cosmology, College of Physical Science and Technology,\\ Yangzhou University, Yangzhou 225009, China}

\author{Xiao-Mei \surname{Kuang}}
\email{xmeikuang@yzu.edu.cn}
\affiliation{Center for Gravitation and Cosmology, College of Physical Science and Technology,\\ Yangzhou University, Yangzhou 225009, China}

\author{Yen Chin \surname{Ong}}
\email{ycong@yzu.edu.cn}
\affiliation{Center for Gravitation and Cosmology, College of Physical Science and Technology,\\ Yangzhou University, Yangzhou 225009, China}
\affiliation{Nordita, KTH Royal Institute of Technology \& Stockholm University, Roslagstullsbacken 23, SE-106 91 Stockholm, Sweden}

\begin{abstract}
The Smarr relation plays an important role in black hole thermodynamics. It is often claimed that the Smarr relation can be written down simply by observing the scaling behavior of the various thermodynamical quantities. We point out that this is not necessarily so in the presence of dimensionful coupling constants, and discuss the issues involving the identification of thermodynamical variables.
\end{abstract}

\maketitle

\section{Smarr Relation and the First Law}

The fact that black holes behave like a thermodynamical system has dramatically changed our understanding of black holes ever since its conception in 1973 \cite{bch}.
For an asymptotically flat Kerr-Newman black hole, the first law of black hole mechanics takes the form
\begin{equation}
\text{d}M= T\text{d}S + \Phi \text{d}Q + \Omega \text{d}J,
\end{equation}
where $M$ denotes the ADM mass of the black hole, $S$ its Bekenstein-Hawking entropy, $T$ its Hawking temperature, $Q$ its electrical charge and $J$ its angular momentum. The first law thus relates the various differential quantities. In some applications, one would like to work directly with the black hole parameters instead of their differentials. Fortunately, there is the Smarr relation \cite{smarr}:
\begin{equation}\label{smarr1}
M=2TS + \Phi Q + 2\Omega J,
\end{equation}
where $\Phi$ denotes the electrical potential, while $\Omega$ denotes the angular velocity of the black hole.

Smarr relations such as this have been widely studied in the literature, beyond the Kerr-Newman family. A good rule of thumb for writing down the Smarr relation for a given black hole is to look at the scaling (i.e. the dimensions) of the various thermodynamical quantities. See, e.g., Sec.2 of \cite{0904.2765}. For example, in 4-dimensions, and in the units $\hbar=k_B=c=1$, we have $M,Q \propto L$, and $J,S\propto L^2$, where $L$ is a length scale. Due to Euler's theorem of quasi-homogeneous function (see below), we can simply write down $M=M(S,Q,J)$ as
\begin{equation}
1\cdot M = 2\cdot \frac{\partial M}{\partial S} S + 1\cdot \frac{\partial M}{\partial Q} Q + 2\cdot \frac{\partial M}{\partial J} J.
\end{equation}

From the first law (and the chain rule), one could identify the various partial derivatives and arrive at the Smarr relation, Eq.(\ref{smarr1}). Similarly, in the extended black hole thermodynamics in which the negative cosmological constant is treated as a thermodynamical variable (a pressure) \cite{1106.6260,1209.1272}, the Smarr relation picks up an additional term $-2(\partial M/\partial \Lambda)\Lambda = -2 VP$, where $V$ is the ``thermodynamical volume''. This identification gives rise to an entire enterprise of ``black hole chemistry'' \cite{1608.06147}.
Note that the coefficient $-2$ comes simply from the dimension of the cosmological constant: $\Lambda \propto L^{-2}$ (which is true in all dimensions). This suggests that if one has a \emph{dimensionless} thermodynamical variable, say $\aleph \propto L^0$, it would not appear in the Smarr relation because the aforementioned rule of thumb would simply yield $0 \cdot (\partial M/\partial \aleph)\aleph =0$.

In this work, we look into the subtleties of such seemingly straightforward statements, and found that contrary to folklore, it \emph{is} possible for the
coefficients that appear in the Smarr relation to differ from the scaling power of the corresponding thermodynamical variables. This is perhaps known to workers in the field, but it is worth emphasizing explicitly, since the counter-example discussed below leads to interesting questions regarding the identification of the thermodynamical variables. Before we get into the issue, it is useful to first review Euler's theorem for quasi-homogeneous functions.

\section{Euler's Theorem for Quasi-Homogeneous Functions}
Let $\left\{\chi^i\right\}_{i=1,\cdots,n}$ be a set of real variables. Let $\left\{\kappa_i\right\}_{i=1,\cdots,n}$ be a set of weights ($\kappa_i \in \Bbb{R}$). A function $\mathcal{F}(\chi^1,\cdots \chi^n): \Bbb{R}^n \mapsto \Bbb{R}$ is called a \emph{quasi-homogeneous equation} of degree $r$ if under re-scaling by a scale factor $\alpha > 0$, one has
\begin{equation}\label{qhe}
\mathcal{F}(\alpha^{\kappa_1}\chi^1,\cdots \alpha^{\kappa_n}\chi^n) = \alpha^r\mathcal{F}(\chi^1,\cdots \chi^n).
\end{equation}
A theorem by Euler states that a differentiable quasi-homogeneous function satisfies
\begin{equation}
\sum_{i=1}^n \kappa_i \chi^i \frac{\partial \mathcal{F}}{\partial \chi^i} = r\mathcal{F}.
\end{equation}
In fact, this is both a sufficient and necessary condition for $\mathcal{F}$ to be quasi-homogeneous \cite{0210031v1, anosov}. If all the weights are equal to 1, then the function is said to be \emph{homogeneous}.

To see how this leads to the Smarr relation, let us consider a simple example: a 4-dimensional asymptotically flat Reissner-Nordstr\"om black hole, whose metric coefficient
\begin{equation}
-g_{tt}=1-\frac{2M}{r}+\frac{Q^2}{r^2},
\end{equation}
yields
\begin{equation}
M=\frac{r_+}{2}\left(1+\frac{Q^2}{r_+^2}\right),
\end{equation}
where $r_+$ denotes the outer (event) horizon.
Equivalently, its ADM mass can be taken as a function of the thermodynamical variables $S=\pi r_+^2$ and $Q$:
\begin{equation}
M(S,Q)=\frac{1}{2}\sqrt{\frac{S}{\pi}}\left(1+\frac{\pi Q^2}{S}\right).
\end{equation}
We note that under re-scaling,
\begin{equation}
M(\alpha^2 S,\alpha Q)=\frac{\alpha}{2}\sqrt{\frac{S}{\pi}}\left(1+\frac{\pi \alpha^2 Q^2}{\alpha^2S}\right) =\alpha M(S,Q).
\end{equation}
Therefore $M(S,Q)$ is a quasi-homogeneous function with degree $r=1$, and Euler's theorem gives the Smarr relation
\begin{equation}
M=2 \frac{\partial M}{\partial S} S + \frac{\partial M}{\partial Q}Q =2 TS + \Phi Q.
\end{equation}
Note that for Smarr relation to hold, $r$ should be a fixed nonzero number in Eq.(\ref{qhe}): there exist known black hole solutions that do not satisfy the standard Smarr relation precisely because this condition is not satisfied. For example, a nonlinear electromagnetic field gave rise to a charged black hole whose metric coefficient is given by
\begin{equation}
-g_{tt} = 1-\frac{2}{r}\left[\frac{Mr^3}{(r^2+q^2)^{\frac{3}{2}}}-\frac{q^2r^3}{2(r^2+q^2)^2}\right],
\end{equation}
where $M$ and $q$ are, respectively, the mass and charge of the black hole.
This is a charged version of the regular Bardeen solution \cite{b1, b2}, in fact the first regular exact black hole solution known \cite{9911046}.
It is straightforward to check that $M(\alpha^x A, \alpha^y q)=\alpha^r M(A,q)$ under the transformation $(A,q) \mapsto (\alpha^x A, \alpha^y q)$ if and only if $x=2y$, and $r=y$, for all $y$, including $y=0$. Indeed, since $A$ has dimension $L^2$ and $q$ has dimension $L$,   if the standard Smarr relation holds, then one should have
\begin{equation}\label{test}
2A\frac{\partial M}{\partial A} + q\frac{\partial M}{\partial q} = M.
\end{equation}
However, explicit calculation shows that the LHS of Eq.(\ref{test}) is identically 0,
which would be absurd since $M$ is the black hole mass. The Smarr relation can nevertheless to be generalized to accommodate black holes coupled with nonlinear electromagnetic fields \cite{1208.6251, 1610.01237, 1710.04660, 1710.07751}.

Let us now proceed to discuss a peculiar case in which the folklore rule does not work, but more interestingly, it leads us to question which quantities are the correct thermodynamical variables.

\section{Smarr Relation and Axionic Coupling}

In \cite{1708.07194}, an asymptotically anti-de Sitter charged flat black hole\footnote{Namely black hole whose horizon has zero sectional curvature. The topology is either a flat torus or planar.} coupled with $k$-essence field was studied in the context of holography.
The action of the theory is
\begin{equation}
I=\int \text{d}^4x \sqrt{-g} \left[\kappa(R-2\Lambda) - \mathcal{K}(X_1,X_2)\right],
\end{equation}
where $\Lambda=-3/l^2$ is the cosmological constant, and the $k$-essence term
\begin{equation}
\mathcal{K}(X_1,X_2)=\sum_{i=1}^2 \left[\frac{1}{2}\nabla^\mu \phi_i \nabla_\mu \phi_i + \gamma\left(\frac{1}{2}\nabla^\mu \phi_i \nabla_\mu \phi_i \right)^k\right]
\end{equation}
is the Lagrangian of two axion fields $\phi_i$, with $X_i:=(1/2)\nabla^\mu \phi_i \nabla_\mu \phi_i$, distributed homogeneously along the coordinates of the planar horizon $x_i$, specifically\footnote{In general, the linear combination of $I$ number of scalar fields take the form $\psi_I=\beta_{Ia}x^a$. We can define
\begin{equation}\notag
\beta^2 := \sum_{a=1}^{n+1}\sum_{I=1}^{n+1}\beta_{Ia}\beta_{Ia}.
\end{equation}
If the coefficients satisfy
\begin{equation}\notag
\sum_{I=1}^{n+1}\beta_{Ia}\beta_{Ib}=\beta^2\delta_{ab},
\end{equation}
then we will obtain the same black hole solution. Since there is rotational symmetry on the underlying space spanned by the coordinates $\{x^a\}$, we can set
$\beta_{Ia}=\beta \delta_{Ia}$ without loss of generality. See related discussions in \cite{1702.01490}.
 } $\phi_i=\lambda x_i$. The action thus consists of a standard kinetic term and a nonlinear one, with a coupling constant $\gamma$, which is dimensionful, usually taken to be positive to avoid phantom instability.

The metric component of the black hole reads (here we ignore the magnetic charge for simplicity)
\begin{equation}\label{k-metric}
-g_{tt} = \frac{r^2}{l^2} - \frac{2M}{r} -\frac{\lambda^2}{2\kappa} + \gamma \frac{\lambda^{2k} r^{2(1-k)}}{2^k(2k-3)\kappa}+\frac{Q^2}{4\kappa r^2}.
\end{equation}
The physical mass and physical charge that satisfy the first laws are: $\mathcal{M} = 4\kappa \sigma M$, $\mathcal{Q}=\sigma Q$. If the horizon is compact (toral), then $\sigma$ is the dimensionless area of the horizon (analogous to $4\pi$ for a 2-sphere). In the planar limit, both $\sigma$ and the physical mass $\mathcal{M}$ tend to infinity, keeping the parameter $M$ finite. Similarly for the electrical charge. That is, $M$ and $Q$ are the mass and charge \emph{density} parameters. The nonlinear power $k$ should be bounded below, $k > 3/2$,  to ensure a proper AdS asymptotic behavior with a well-defined mass \cite{1708.07194}.

The authors of  \cite{1708.07194} emphasized the importance of treating $\lambda$ as the axionic charge when investigating phase transitions, as well as in the holographic contexts. In fact the linear term and the nonlinear terms in the $k$-essence Lagrangian $\mathcal{K}(X_1,X_2)$ gave rise to two physical axionic charges: $\mathcal{Q}_i = -\sigma \lambda$ and $\mathcal{Q}_{i,k}=-\sigma \gamma \lambda^k$. More specifically, the thermodynamic analysis was carried out via Euclidean formalism in \cite{1708.07194}, in which the Euclidean action is related to the Gibbs free energy $\mathcal{G}$, which in turn is equivalent to Euclidean Hamiltonian
action evaluated on-shell, by
\begin{flalign}\notag
I_E&=\beta \mathcal{G} \\ \notag
&= \beta \mathcal{M} - S - \beta A(r_h) \mathcal{Q} - \beta \sum_{i=1}^2 \left(\hat\Psi_{i}\mathcal{Q}_i + \hat\Psi_{i,k}\mathcal{Q}_{i,k}\right), \notag
\end{flalign}
where $\beta$ is the inverse temperature of the black hole,   $\hat{\Psi}_{i}:={\partial \mathcal{M}}/{\partial \mathcal{Q}_{i}}$, and $\hat{\Psi}_{i,k}:={\partial \mathcal{M}}/{\partial \mathcal{Q}_{i,k}}$.
Following standard procedures, one can obtain the conserved charges $\mathcal{M}$, $\mathcal{Q}$, $\mathcal{Q}_i$, and $\mathcal{Q}_{i,k}$ essentially via taking the derivatives of the action. In particular, the axionic charges are
\begin{equation}\notag
\mathcal{Q}_i = -\frac{1}{\beta}\left(\frac{\partial I_E}{\partial \hat\Psi_{i}}\right)_\beta = \sigma \lambda,
\end{equation}
and
\begin{equation}\notag
\mathcal{Q}_{i,k}=-\frac{1}{\beta}\left(\frac{\partial I_E}{\partial \hat\Psi_{i,k}}\right)_\beta = -\sigma \gamma \lambda^k.
\end{equation}

We can express the mass in terms of other black hole parameters:
\begin{flalign}\label{eqm}
&\mathcal{M}(\mathcal{A}, \Lambda, \mathcal{Q}, \mathcal{Q}_i, \mathcal{Q}_{i,k}) \\ \notag
&=2\sqrt{\sigma \mathcal{A}} \left[-\frac{\mathcal{A}\Lambda}{3\sigma}-\frac{\mathcal{Q}_i^2}{2\sigma^2} + \frac{\mathcal{Q}_{i,k}^2}{\sigma^2 \gamma 2^k (2k-3)}\left(\frac{\mathcal{A}}{\sigma}\right)^{1-k} \right. \\
&\quad\left.+\frac{\mathcal{Q}^2}{4\sigma \mathcal{A}} \right],\notag
\end{flalign}
where $\mathcal{A}=\sigma r_+^2$ is the area of the event horizon. One notes that this expression of $\mathcal{M}$ contains a dimensionful parameter $\gamma$, which \emph{cannot} be eliminated. This is because $\gamma$ only occurs alongside $\lambda^k$, while $\lambda^k$ and $\lambda$ (or more precisely, $\mathcal{Q}_i$ and $\mathcal{Q}_{i,k}$) are being treated as two \emph{distinct} thermodynamical variables. In other words, $\mathcal{M}$ cannot be fully expressed in terms of the thermodynamical variables $\mathcal{A}, \Lambda, \mathcal{Q}, \mathcal{Q}_i, \mathcal{Q}_{i,k}$. \emph{The presence of this dimensionful coupling constant is the reason behind the failure of the standard ``folklore'' method as we will see below}.

Under the scaling transformation
\begin{equation}
(\mathcal{A}, \Lambda, \mathcal{Q}, \mathcal{Q}_i, \mathcal{Q}_{i,k}) \mapsto (\alpha^2\mathcal{A}, \alpha^{-2}\Lambda, \alpha \mathcal{Q}, \alpha^x\mathcal{Q}_i, \alpha^y\mathcal{Q}_{i,k}), \notag
\end{equation}
one can check that $\mathcal{M}$ is homogeneous, i.e.,
\begin{flalign}
&\mathcal{M}(\alpha^2\mathcal{A}, \alpha^{-2}\Lambda, \alpha \mathcal{Q}, \alpha^x\mathcal{Q}_i, \alpha^y\mathcal{Q}_{i,k}) \notag \\
& = \alpha \mathcal{M}(\mathcal{A}, \Lambda, \mathcal{Q}, \mathcal{Q}_i, \mathcal{Q}_{i,k}), \notag
\end{flalign}
if and only if $x=0$ and $y=k-1$. The fact that $x=0$ is expected since $\mathcal{Q}_i$ is dimensionless. However, $\mathcal{Q}_{i,k}$ has the same dimension as $\gamma$, which from Eq.(\ref{k-metric}), can be obtained to be $L^{2k-2}$. That is, $\text{dim}(\mathcal{Q}_{i,k})=L^{2k-2} \neq L^y$. Consequently, the Smarr relation has a term (for each $i$),
\begin{equation}
(k-1)\mathcal{Q}_{i,k} \frac{\partial \mathcal{M}}{\partial \mathcal{Q}_{i,k}} \subset \mathcal{M}.
\end{equation}
If one naively writes down the Smarr relation following the folklore, one would have written down a wrong term
\begin{equation}
(2k-2)\mathcal{Q}_{i,k} \frac{\partial \mathcal{M}}{\partial \mathcal{Q}_{i,k}}
\end{equation}
instead, which is twice as large.

The correct Smarr relation is, in full,
\begin{equation}\notag
\mathcal{M}=2TS + \Phi\mathcal{Q}-2PV + (k-1)\sum_{i=1}^2\left(\hat{\Psi}_{i,k}\mathcal{Q}_{i,k}\right).
\end{equation}

The reason for this is that a dimensionful coupling constant $\gamma$ appears in the same term as $\mathcal{Q}_{i,k}$ in  Eq.(\ref{eqm}), but $\gamma$ itself is not a thermodynamical variable so it does not contribute to power of $\alpha$ under the scaling transformation, neither can it be eliminated from the expression of the mass as we have previously mentioned. It is not easy to know this a priori from the metric tensor Eq.(\ref{k-metric}) since Smarr relation requires one to first identify the correct physical quantities that appear in the first law of black hole mechanics (mistaking a mere mass parameter as the true physical mass do sometimes happen, which would lead to incorrect conclusion when dealing with physical processes \cite{1506.01248}). Such a phenomenon does not occur in, say, AdS-Kerr black hole, despite its physical mass $E:=M/(1-a^2/l^2)^2$ involves two other dimensionful parameters: the rotation parameter $a$ and cosmological constant length scale $l$. Whether or not $a$ and $l$ are thermodynamical variables are irrelevant here, since their dimensions cancel in taking the ratio.

Since $\gamma$ only appears together with $\mathcal{Q}_{i,k}$, an alternative interpretation would be to take $\gamma$ as the thermodynamical variable, instead of $\mathcal{Q}_{i,k}$ \cite{1808.00176}. We would then obtain a Smarr relation in which the ``folklore method'' \emph{does} lead to the correct form
\begin{equation}\label{smarr2}
\mathcal{M}=2 T S + \Phi  \mathcal{Q} -2PV  + (2k-2)\varphi \gamma,
\end{equation}
with
$
\varphi := {\partial \mathcal{M}}/{\partial \gamma}.
$

\section{Discussion}
In this short note we point out that the usual ``folklore method'', which allows one to write down the Smarr relation of a black hole by merely inspecting the dimension of the thermodynamical variables, may fail if the terms contain dimensionful coupling constants (which are not thermodynamical variable). This means that one has to exercise extra caution in the presence of such constants. We provide a concrete example to illustrate this: a charged AdS flat black hole coupled to $k$-essence fields. (A similar issue which involves dimensionful coupling constant was investigated in the context of Lovelock gravity in \cite{1005.5053}.)

As we recall, $\mathcal{Q}_i=-\sigma \lambda$ and $\mathcal{Q}_{i,k}=-\sigma \gamma \lambda^k$.
For $\mathcal{Q}_i$ it is clear that the conserved charge is, modulo the dimensionless area, the same as $\lambda$.
It is possible to ``save'' the folklore method by re-interpreting the coupling constant $\gamma$ as a thermodynamical variable. In this interpretation, we can consider varying the coupling ``constant'', much like varying the cosmological ``constant'' to produce the pressure term.
Note that both axionic charges appear in the first law of the axionic charged black hole (see \cite{1708.07194}, though the pressure term is not considered therein)
\begin{equation}\label{dM1}
\text{d}\mathcal{M}=T \text{d}S+ \Phi \text{d}\mathcal{Q} + V\text{d}P + \sum_{i=1}^2\left[\hat\Psi_i \text{d}\mathcal{Q}_i + \hat\Psi_{i,k} \text{d}\mathcal{Q}_{i,k}\right],
\end{equation}
where $\hat\Psi_i:=\partial \mathcal{M}/\partial \mathcal{Q}_i$, and $\hat\Psi_{i,k}:=\partial \mathcal{M}/\partial \mathcal{Q}_{i,k}$.

Thus if we treat $\gamma$ as thermodynamical variable in the $\mathcal{Q}_{i,k}$ term instead of $\lambda^k$, we have -- equivalently --
\begin{equation}\label{dM2}
\text{d}\mathcal{M}=T \text{d}S+ \Phi \text{d}\mathcal{Q} + V\text{d}P +  \Xi \text{d}\lambda + \varphi \text{d}\gamma,
\end{equation}
where $\Xi:=\partial{\mathcal{M}}/\partial{\lambda}$. Note that since the axion fields are homogeneous and isotropic, we treat $\mathcal{Q}_{i}$ as identical for both $i=1,2$. The same goes for $\mathcal{Q}_{i,k}$ Subsequently, the sum of quantities for two axions fields in Eq.\eqref{dM1} does not appear in Eq.\eqref{dM2}. 
Since $\mathcal{Q}_i$ and $\mathcal{Q}_{i,k}$ are distinct charges, it makes sense that $\lambda$ and $\gamma$ should be treated independently as thermodynamical variables.

However, if one interprets instead that $\lambda^k$ arises from $\lambda$, then one might want to argue that $\lambda$ no longer plays any role as thermodynamical variable once $\lambda^k$ does not. In this case the $\mathcal{Q}_i$ term is absent in the first law (since $\mathcal{Q}_i$ is not accompanied by a coupling constant, or more precisely, one may say that its coupling constant is unity, which unlike $\gamma$, cannot be promoted to be a parameter), as was considered in \cite{1808.00176}. It can be shown that both first laws are valid if consistently derived, and the Smarr relation Eq.(\ref{smarr2}) is valid for both of them (since $\mathcal{Q}_i$ term does not contribute even when present in the first law).

Of course, mathematically both interpretations are consistent. Physically however, one must ask: does the validity of the ``folklore method'' actually give us more insight into the \emph{physics} (of which parameters are better chosen as the ``right'' thermodynamical variables)? We leave this puzzle open for future investigations.


\begin{acknowledgments}
XMK is supported by the Natural Science Foundation of China (grant No.11705161) and Natural Science Foundation of Jiangsu Province (grant No.BK20170481).
YCO is supported by the National Natural Science Foundation of China (grant No.11705162) and the Natural Science Foundation of Jiangsu Province (No.BK20170479).
YCO thanks Nordita, where part of this work was carried out, for hospitality during his summer visit and participation in the Lambda program.
The authors also thank all members of Center for Gravitation and Cosmology (CGC) of Yangzhou University (\href{http://www.cgc-yzu.cn}{http://www.cgc-yzu.cn}) for discussions and various supports.
\end{acknowledgments}

\end{document}